\documentclass[runningheads]{casc} 

\def\@algheader{{\bf Algorithm: }}
\def\@alginheader{Input:}
\def\@algoutheader{Output:}

\usepackage{ifthen}
\def\JGDeg{\kern-.14em\lower-.3ex\hbox{\ensuremath{{}^{\circ}\,}}\hspace{-.2em}}

\renewcommand{\lg}{\mathop{\operator@font log}\nolimits}
\renewcommand{\log}[1][e]{ \ifthenelse{\equal{#1}{e}}
  {\mathop{\operator@font ln}\nolimits} {\mathop{\operator@font
      log_{#1}}\nolimits}}

\def\sO{{\mathcal O}\lower 2pt\hbox{\large\ensuremath{{}^{\widetilde{}}}}}

\DeclareRobustCommand\sfrac[1]{\@ifnextchar/{\@sfrac{#1}}%
                                            {\@sfrac{#1}/}}
\DeclareRobustCommand\jfrac[1]{\@ifnextchar/{\@jfrac{#1}}%
                                            {\@jfrac{#1}/}}
\def\@sfrac#1/#2{\leavevmode\kern.1em\raise.0ex
         \hbox{$\m@th\mbox{#1}$}\kern-.1em
         /\kern-.15em\lower.3ex
          \hbox{$\m@th\mbox{#2}$}}

\def\@jfrac#1/#2#3{\leavevmode\kern.1em\raise.0ex
         \hbox{$\m@th\mbox{#1}$}\kern-.1em
         /\kern-.15em\lower.3ex
          \hbox{$\m@th\mbox{\fontsize\sf@size\z@
                            \selectfont#2}\m@th\mbox{#3}$}}

\newcommand{\Z}{\ensuremath{\mathbb Z}}

\newcommand{\pF}[1][vide]{\ifthenelse{\equal{#1}{vide}}{\Z}{\leavevmode
	\kern.1em\raise.0ex \hbox{\Z}\kern-.1em /\kern-.15em\lower.3ex
         \hbox{#1}\mbox{\Z}}
}
\newcommand{\po}{p_{\kern-.3em\hbox{\ensuremath{{}_{{}_{0}}}}}\kern-0.06em}

\newcommand{\GF}[1][vide]{\ifthenelse{\equal{#1}{vide}}{}{\ensuremath{\mathtt {GF}(#1)}}}

\newlength{\Stmtindent}
\makeatletter

\newcommand{\til}{\lower 2pt\hbox{\small${}^\sim$}}

\usepackage{graphicx}
\usepackage{amssymb,amsfonts,amsmath}

\title{Efficient dot product over word-size finite fields}
\titlerunning{Finite field dot product}
\author{Jean-Guillaume Dumas} 
\authorrunning{Jean-Guillaume Dumas}
\institute{Laboratoire de Mod\'elisation et Calcul.  50 av. des
  Math\'ematiques, B.P. 53, 38041 Grenoble, France.
\email{Jean-Guillaume.Dumas@imag.fr, www-lmc.imag.fr/lmc-mosaic/Jean-Guillaume.Dumas}
}
\begin{document}
\pagestyle{headings}
\mainmatter
\maketitle

\begin{abstract}
We want to achieve efficiency
  for the exact computation  of
  the dot product of two vectors over word size finite fields. 
We therefore compare the practical behaviors of a wide range of
implementation techniques using different representations. The
techniques used include floating point representations, discrete
logarithms, tabulations, Montgomery reduction, delayed modulus.

\end{abstract}

\section{Introduction}\label{sec:intro}

Dot product is the core of many linear algebra routines. Fast routines
for matrix
multiplication \cite[\S 3.3.3]{jgd:2002:issac}, triangular system
solving and matrix factorizations \cite[\S 3.3]{jgd:2004:issac} on
the one hand, iterative methods on the other hand (see
e.g. \cite{Wiedemann:1986:SSLE,LaMacchia:1991:SLS,jgd:2002:villard}, etc.)
make extensive usage of an efficient dot product over finite fields.
In this paper, we compare the practical behavior of many implementations
of this essential routine. Of course, this behavior is highly dependent on the
machine arithmetic and on the finite field representations used. 
Section \ref{sec:rep} therefore proposes several possible
representations while section~\ref{sec:dp} presents some algorithms
for the dot product itself. Experiments and choice of
representation/algorithm are also presented in section~\ref{sec:dp}.

\section{Prime field representations}
\label{sec:rep}
We present here various methods implementing seven of the basic
arithmetic operations: the addition, the subtraction, the negation,
the multiplication, the division, a multiplication followed by an
addition ($r \leftarrow  a*x+y$) or {\em AXPY} (also called
``fused-mac'' within hardware) and a multiplication followed
by an in-place addition ($r  \leftarrow a*x + r$) or {\em AXPYIN}. 
Within linear algebra in general (e.g. Gaussian elimination) 
and for dot product in particular, these last two operations are the
most widely used. We now present different ways to implement these
operations. The compiler used for the C/C++ programs was ``gcc version 3.2.3 20030309 (Debian prerelease)''.

\subsection{Classical representation with integer division}\label{sec:zpz}
{\em Zpz} is the classical representation, with positive values
between 0 and $p-1$, for $p$ the prime.
\begin{itemize}
\item Addition is a signed addition followed by a test. An additional
    subtraction of the modulus is made when necessary. 
\item Subtraction is similar.
\item Multiplication is machine multiplication and machine remainder.
\item Division is performed via the extended gcd algorithm. 
\item {\em AXPY} is a machine multiplication and a machine addition
  followed by only one machine remainder.
\end{itemize}
For the results to be correct, the intermediate AXPY value must not
overflow. For a $m-$bit machine integer, the prime must therefore 
be below $2^{\frac{m-1}{2}}-1$ if signed values are used. For 32 and
64 bits this gives primes below $46337$ and below $3037000493$.\\

Note that with some care those operations can be extended to work with
unsigned machine integers. The possible primes then 
being below $65521$ and $4294967291$.

\subsection{Montgomery representation}\label{ssec:montg}
To avoid the costly machine remainders, Montgomery designed another
reduction \cite{Montgomery:1985:MMT}: when $gcd(p,B)=1$,
$n_{im} \equiv -p^{-1} \mod B$ and $T$ is such that $0 \leq T \leq pB$,
if $U \equiv T n_{im} \mod B$, then $(T+Up)/B$ is an integer and
$(T+Up)/B \equiv TB^{-1} \mod p$. The idea of Montgomery is to set
$B$ to half the word size so that multiplication and divisions by $B$
will just be shifts and remaindering by $B$ is just application of a
bit-mask. Then, one can use the reduction to perform the
remainderings by $p$. Indeed one shift, two bit-masks and two machine
multiplications are very often much less expensive than a machine
remaindering. \\

The idea is then to change the representation of the elements: every
element $a$ is stored as $aB \mod p$. Then additions, subtractions
are unchanged and the prime field multiplication is now a
machine multiplication, followed by only one Montgomery reduction.
Nevertheless, One has to be careful when implementing the AXPY
operator since $axB^2$ cannot be added to $yB$ directly. 
Then
the primes must verify $(p-1)^2 + p*(B-1) < B^2$, which gives 
$p \leq 40499$ (resp. $p \leq 2654435761$) for $B=2^{16}$ (resp. $B=2^{32}$).

\subsection{Floating point representation}
Yet another way to perform the reduction is to use the floating
point routines. According to \cite[Table 3.1]{Defour:2003:these}, for
most of the architectures (alpha, amd, Pentium IV, Itanium, sun, etc.)
those routines are faster than the integer ones (except for the
Pentium III). The idea is then to compute $T \mod  p$ by way of a
precomputation of a high precision numerical inverse of $p$:
$T \mod  p = T - \lfloor T * \frac{1}{p}  \rfloor * p$. \\

The idea here
is that floating point division and truncation are quite fast when
compared to machine remaindering. 
Now on floating point architectures the round-off can induce a $\pm 1$
error when the flooring is computed. This requires then an adjustment
as implemented e.g. in Shoup's NTL \cite{Shoup:NTL}~:

\medskip\par\noindent\begin{minipage}{13cm}
\hrule\vspace{3pt}
NTL's floating point reduction
\vspace{3pt}\hrule\vspace{2pt}\begin{verbatim}
  double P, invP, T; 
  ...
  T  -= floor(T*invP)*P;
  if (T >= P)     T -= P;
  else if (T < 0) T += P;
\end{verbatim}\hrule\end{minipage}

\subsection{Discrete logarithms}\label{sec:gfq}
This representation is also known as Zech logarithms, see
e.g. \cite{Douillet:2001:zech} and references therein.
The idea is to use a generator of the multiplicative group, namely a
primitive element. Then, every non zero element is a power of this
primitive element and this exponent can be used as an internal
representation:
\[ \begin{cases} 0 & \text{if } x = 0 \\
q-1 & \text{if } x = 1 \\
i & \text{if } x = g^i \text{and } 1 \leq i < q-1 \end{cases} \]
Then many tricks can be used to perform the operations that require
some extra tables see e.g. \cite{Huber:1990:Zech,Bailey:2001:EAFFE}.
This representation can be used for prime fields as well as for their
extensions, we will therefore use the notation $\GF[q]$.
The operations are then:
\begin{itemize} 
\item Multiplication and division of invertible elements are just an
  index addition and subtraction modulo $\overline{q} = q-1$. 
\item Negation is identity in characteristic $2$ and addition of
  $i_{-1} = \frac{q-1}{2}$ modulo $\overline{q} =$ in odd characteristic.
\item Addition is now quite complex. If $g^i$ and $g^j$ are invertibles
  to be added then their sum is $g^i + g^j = g^i(1+g^{j-i})$ which can
  be implemented using index addition and subtraction and access to a
  ``plus one'' table ($t\_plus1[]$) of size $q$ giving the exponent
  $h$ of any number of the form $1+g^k$, so that $g^h=1+g^k$.
\end{itemize}

\begin{table}[ht]
\centering
\begin{tabular}{|l||c|c|c|c|c|}
\hline
Operation & \hspace{1em}Elements\hspace{1em} & \hspace{3em}Indices\hspace{3em} & \multicolumn{3}{c|}{Cost}\\
    & & & \hspace{2em}+/-\hspace{2em} & \hspace{1em}Tests\hspace{1em} & Accesses \\
\hline
\vspace{-1ex}& & & & &\\
Multiplication\hspace{1em} & $g^i * g^j$ & $i+j\ (- \overline{q})$ & 1.5 & 1 & 0\\[1ex]
Division & $g^i / g^j$ & $ i - j\ (+ \overline{q})$ & 1.5 & 1 & 0\\[1ex]
Negation & $-g^i$ & $i-i_{-1}\ (+ \overline{q})$& 1.5 & 1 & 0 \\[1ex]
Addition & $g^i + g^j$ & $k = j-i\ (+ \overline{q})$ &  & & \\
 &  & $i+t\_plus1[k]\ (- \overline{q})$ & 3 & 2 & 1\\[1ex]
Subtraction & $g^i - g^j$ & $k = j-i+i_{-1}\ (\pm \overline{q})$ &
 & & \\
 &  & $i+t\_plus1[k]\ (- \overline{q})$ &
3.75 & 2.875 & 1 \\
\hline
\end{tabular}
\caption{Number of elementary operations to implement Zech logarithms
  for an odd characteristic}\label{tab:oper} 
\end{table}

Table \ref{tab:oper} shows the number of elementary operations to
implement Zech logarithms for an odd characteristic finite field.
Only one table of size $q$ is considered. This number is divided into
three types: mean number of exponent additions and subtractions
(+/-), number of tests and number of table accesses. \\

\begin{figure}[htbp]\begin{center}
\includegraphics{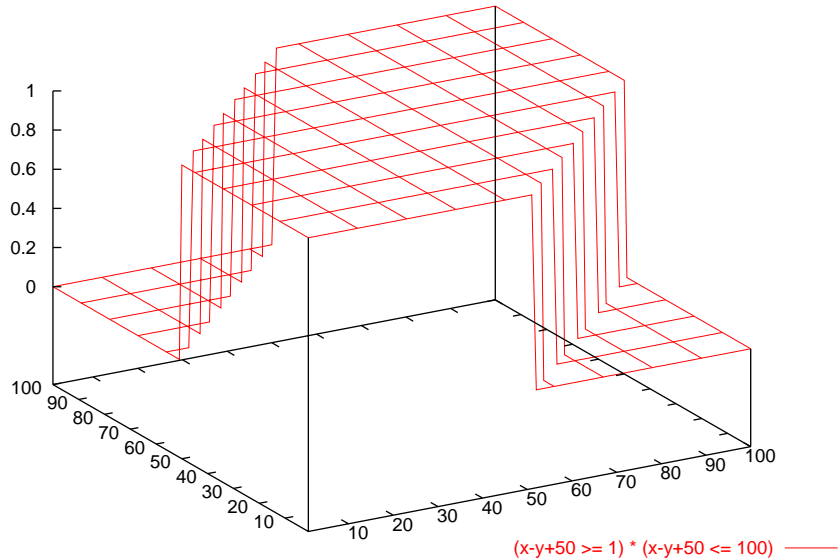}
\caption{$j-i+\frac{q-1}{2}$ for $i$ and $j$ between $1$ and $q-1$, for
  $q=101$}\label{fig:soustraction}
\end{center}\end{figure}

We have counted $1.5$ index operations when a correction by
$\overline{q}$ actually arises only for half the possible values.
The fact that the
mean number of index operations is 3.75 for the subtraction is easily
proved in view of the possible values taken by $j-i+\frac{q-1}{2}$ for
varying $i$ and $j$. 
In this case, $j-i+i_{-1}$ is between $-\frac{\overline{q}}{2}$ and
$\frac{3\overline{q}}{2}$ and requires a correction by $\overline{q}$
only two eighth of the time as shown in figure \ref{fig:soustraction} for
$q=101$. 
The total number of additions or subtractions is then 
$2+0.25+1+0.5=3.75$ and the number of tests  $1+0.875+1=2.875$ follows
(one test towards zero, one only in case of a positive value
i.e. seven eighth of the time, and a last one after the table lookup). 
It is possible to actually reduce the number of index exponents,
(for instance replacing $i+x$ by $i+x-\frac{q-1}{2}$) but to the price
of an extra test. In general, such a test ($a > q~?$) is as costly as
the $a-q$ operation. We therefore propose an implementation minimizing
the total cost with a single table.\\

These operations are valid as long as the index operations do not
overflow, since those are just signed additions or subtractions.
This gives maximal prime value of e.g. $1073741789$ for $32$ bits
integer. However, the table size is now a limiting factor: indeed with
a field of size $2^{26}$ the table is already $256$ Mb. Table
reduction is then mandatory. We will not deal with such optimizations
in this paper, see e.g. \cite{Huber:1992:gfpm,Douillet:2001:zech} for
more details.

\subsubsection{Fully tabulated}\label{sec:gfqtab}
A last possibility is to further tabulate the Zech logarithm
representation. The idea is to code $0$ by $2 \overline{q}$ instead of
$0$. Then a table can be made for the multiplication:
\begin{itemize} 
\item $t\_mul[k] = k$ for $0 \leq k < \overline{q}$.
\item $t\_mul[k] = k-\overline{q}$ for $q-1 \leq k < 2 \overline{q}$.
\item $t\_mul[k] = 2 \overline{q}$ 
for $2 \overline{q} \leq k \leq 4 \overline{q}$.
\end{itemize} 
The same can be done for the division via a shift of the table and
creation of the negative values, thus giving a table of size $5q$.
For the addition, the $t\_plus1$ has also to be extended to other
values, to a size of $4q$. For subtraction, an extra table of size
$4q$ has also to be created. When adding the back and forth conversion
tables, this gives a total of $15q$. This becomes quite huge and quite
useless nowadays when memory accesses are a lot more expensive
than arithmetic operations.
\subsubsection{Field extensions}\label{sec:ext} 
Another very interesting property is that whenever this implementation
is not at all valid for non prime modulus, it remains identical for
field extensions. In this case the classical representation would
introduce polynomial arithmetic. This discrete logarithm
representation, on the contrary, would remain atomic, thus inducing a
speed-up factor of $O(d^2)$, for $d$ the extension degree. See
e.g. \cite[\S 4]{jgd:2002:issac} for more details.

\subsection{Atomic comparisons}
\begin{figure}[htbp]\begin{center}
\includegraphics[angle=90,width=\textwidth]{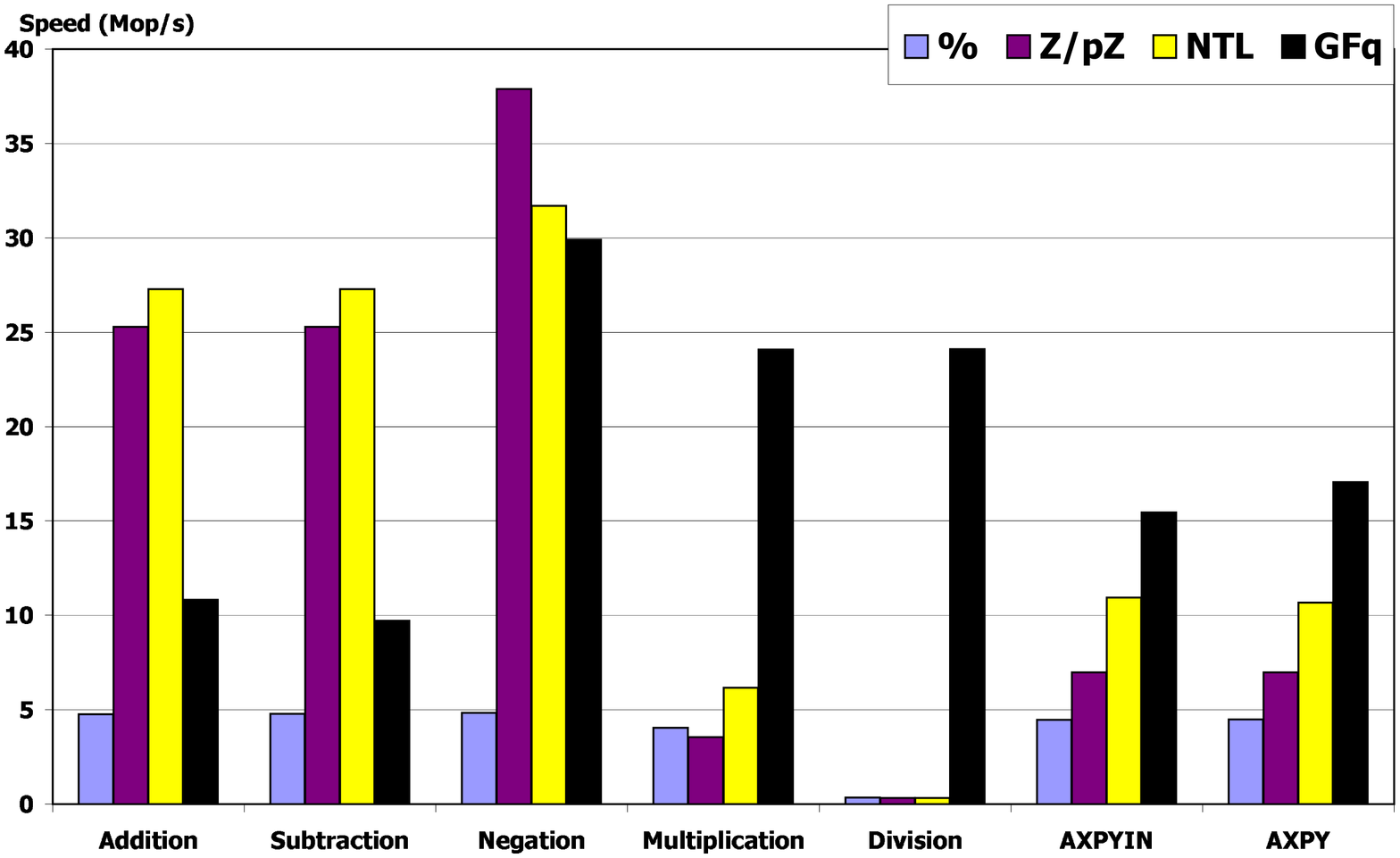}
\caption{Single arithmetic operation modulo 32749 on a sparc ultra II, 250 MHz}
\label{fig:arith32749}
\end{center}\end{figure}

\begin{figure}[htbp]\begin{center}
\includegraphics[angle=90,width=\textwidth]{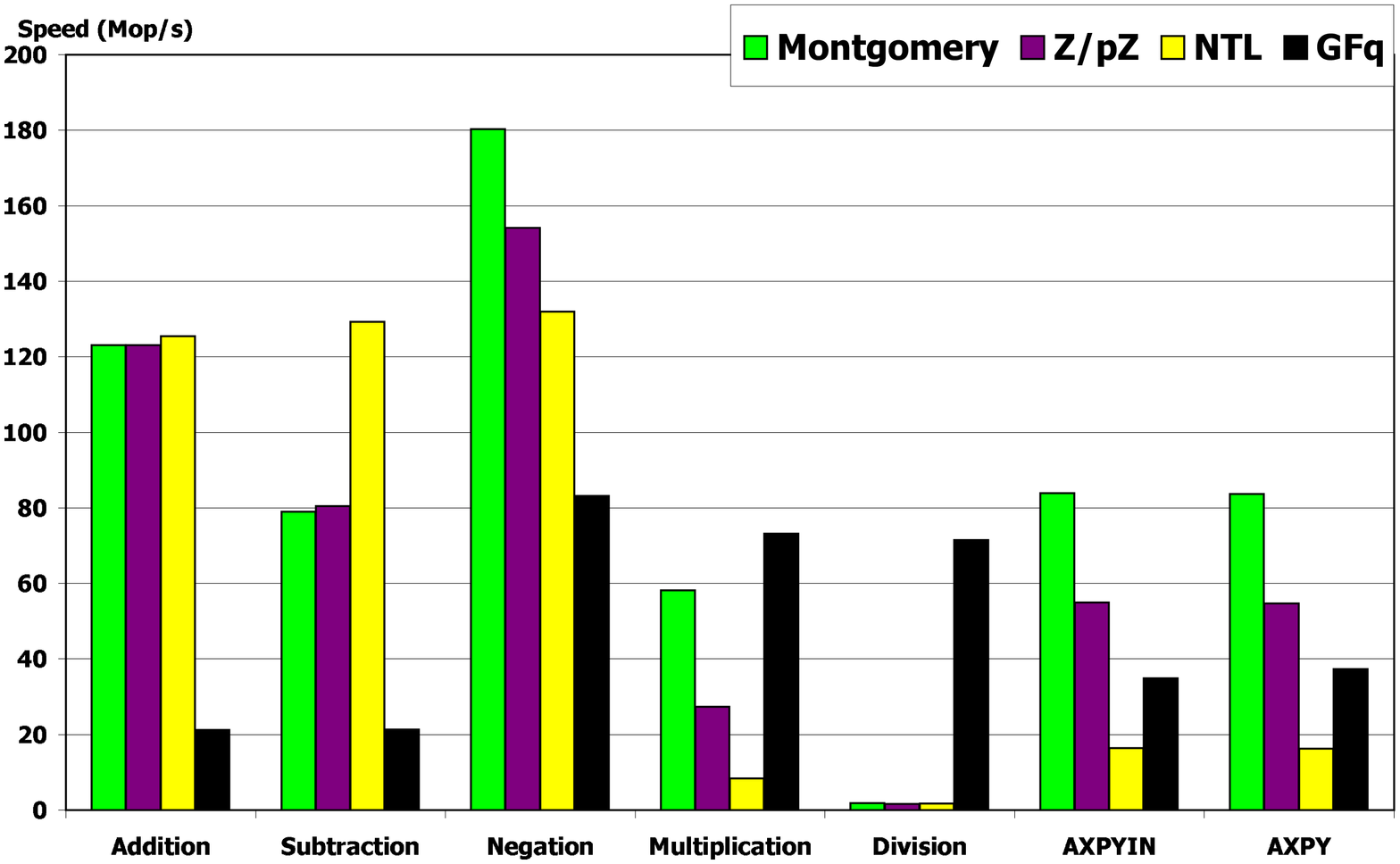}
\caption{Single arithmetic operation modulo 32749 on a Pentium III, 1 GHz}
\label{fig:arithP3}
\end{center}\end{figure}

\begin{figure}[htbp]\begin{center}
\includegraphics[angle=90,width=\textwidth]{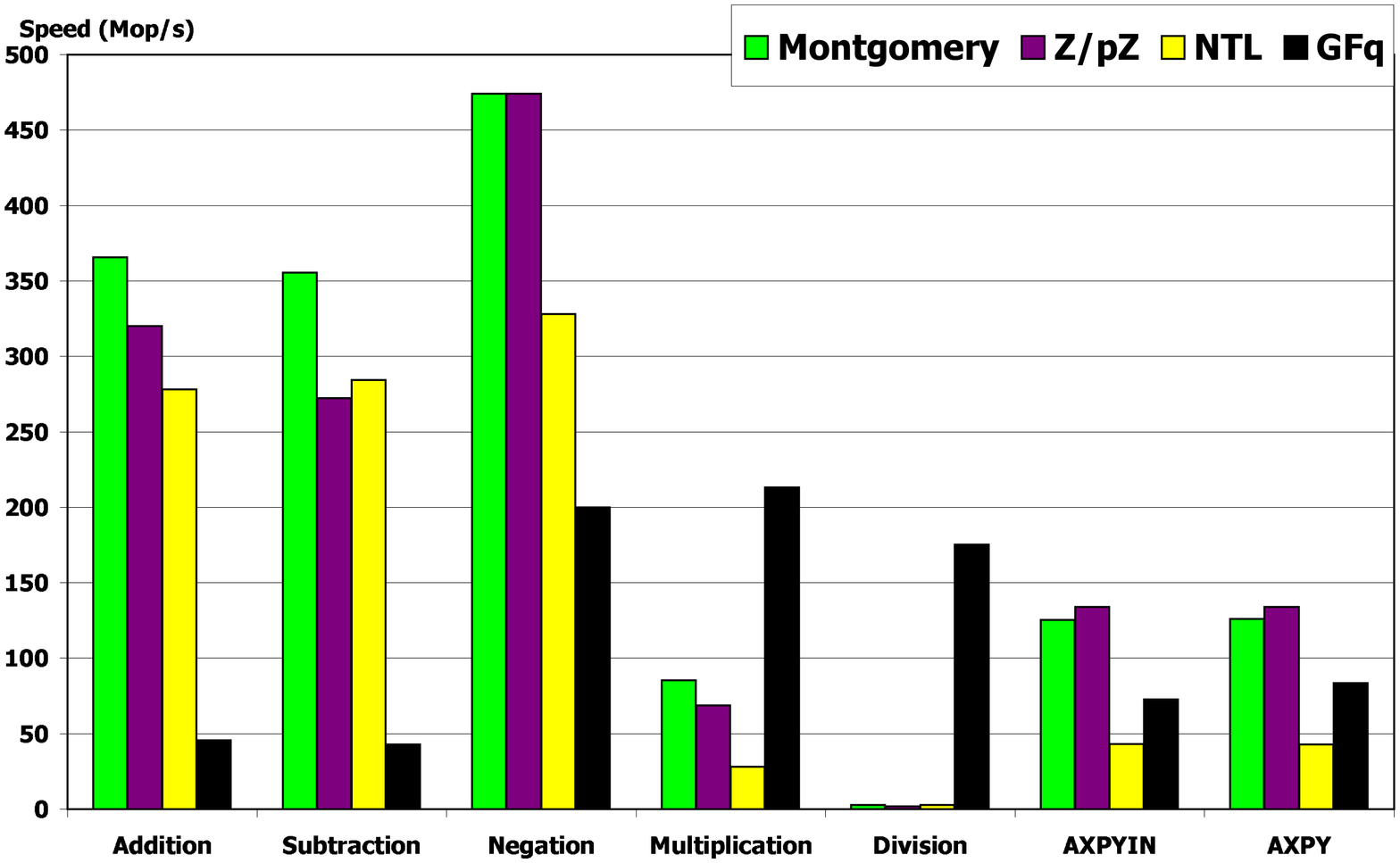}
\caption{Single arithmetic operation modulo 32749 on a Pentium IV, 2.4 GHz}
\label{fig:arithP4}
\end{center}\end{figure}

We first present a comparison between the preceding implementation
possibilities. The idea is to compare just the atomic operations.
``\%'' denotes an implementation using machine remaindering (machine division)
 for every operation. This is just to give a comparing scale.
``NTL'' denotes NTL's floating point flooring for multiplication ;
``Z/pZ'' denotes our implementation of the classical representation
when tests ensure that machine remaindering is used only when really
needed. Last ``GFq'' denotes the discrete logarithm implementation of
section \ref{sec:gfq}. In order to be able to compare those single
operations, the experiment is an application of the arithmetic
operator on vectors of a given size (e.g. 256 for figures
\ref{fig:arith32749}, \ref{fig:arithP3} and \ref{fig:arithP4}).
%

We compare the number of millions of field arithmetic operations per
second, {\em Mop/s}.\\

Figure \ref{fig:arith32749} shows the results on a UltraSparc II 250
Mhz. 
First one can see that the need of Euclid's algorithm for the field
division is a huge
drawback of all the implementations save one. Indeed division over
``GFq'' are just an index subtraction.
Next, we see that floating point operations are quite
faster than integer operations: indeed, NTL's multiplication is better
than the integer one.
Now, on this machine, memory accesses are not yet much slower than
arithmetic operations. This is the reason why discrete logarithm
addition is only 2 to 3 times slower than one arithmetic call. This,
enables the ``GFq'' AXPY (base operation of most of the linear algebra
operators) to be the fastest.\\

On newer PC, namely Intel Pentium III and IV of figures \ref{fig:arithP3}
and \ref{fig:arithP4}, one can see that now memory accesses have
become much too slow. Then any tabulated implementation is penalized,
except for extremely small modulus. NTL's implementation is also
penalized, both because of better integer operations and because of a
pretty bad flooring (casting to integer) of the floating point
representation.
Now, for Montgomery reduction, this trick is very efficient for the
multiplication. However; it seems that it becomes less useful as the
machine division improves as shown by the AXPY results of
figure \ref{fig:arithP4}. As shown section \ref{ssec:montg} the
Montgomery AXPY is less impressive because one has to compute first the
multiplication, one reduction and then only the addition and
tests. This is due to our choice of representation $aB$. ``Zpz'', not
suffering from this distinction between multiplied and added values
can perform the multiplication and addition before the reduction so
that one test and sometimes a correction by $p$ are
saved. Nevertheless, we will see in next section that our choice of
representation is not anymore a disadvantage for the dot product.
\section{Dot products}\label{sec:dp}
In this section, we extend the results of \cite[\S
3.1]{jgd:2002:issac}. To main techniques are used: regrouping
operations before performing the remaindering, and performing this
remaindering only on demand. Several new variants of the representations
of section \ref{sec:rep} are tested and compared. For ``GFq'' and
``Montgomery'' representations the dot products are of course
performed with their representations. In particular the timings
presented do not include conversions. The argument is that the
dot product is computed to be used within another higher
computation. Therefore the conversion will only be useful for reading
the values in the beginning and for writing the result at the end of
the whole program.

\subsection{53 and 64 bits}
\begin{figure}[htb]\begin{center}
\includegraphics[width=\columnwidth]{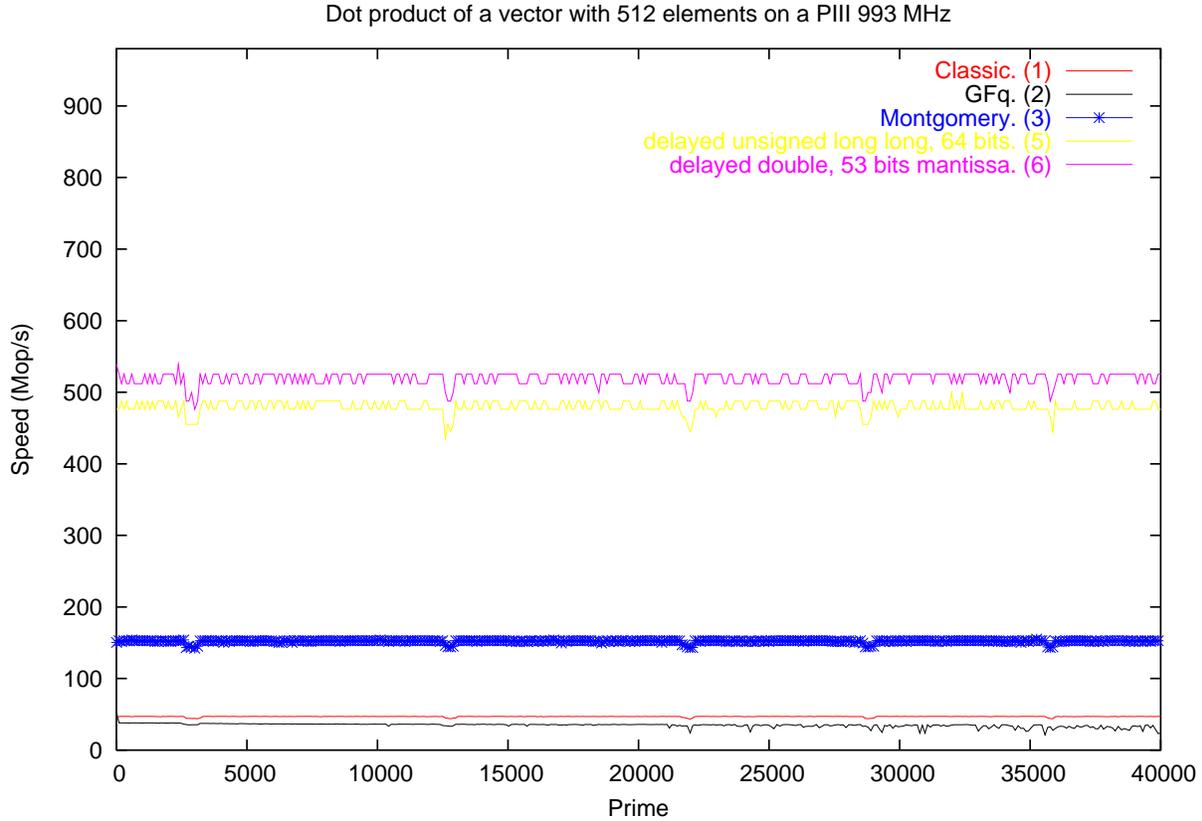}
\caption{Dot product by delayed division, on a PIII}
\label{fig:dotprod_F5}
\end{center}\end{figure}

The first idea is to use a representation where no overflow can occur
during the course of the dot product so that the division is delayed
to the end of the product:
if the available mantissa is of $m$
bits and the modulo is $p$, the division happens at worst every $\lambda$
multiplications where $\lambda$ verifies the following condition:
\begin{equation}\label{eq:fond}
  \lambda (p-1)^2 < 2^m
\end{equation}
For instance when using 64 bits
integers with numbers modulo 40009, a dot product of vectors of size
lower than $1.15 10^{10}$ will never overflow. Therefore one has just
to perform the AXPY without any division. A single machine
remaindering is needed at the end for the whole computation.
This produces very good speed ups for $53$ (double
representation) and $64$ bits storage as shown in figure
\ref{fig:dotprod_F5}. There, the floating point representation performs the division ``\`a la NTL'' using a floating point precomputation of
the inverse, so that it is slightly better than the $64-$bit integer
representation. 
Note also the very good behavior of an implementation of P. Zimmermann
\cite{Zimmermann:2002:personalcommunication} of Montgomery reduction
over $32-$bit integers. 

\subsection{AXPY blocks}\label{ssec:del}
The extension of this idea is to specialize dot product in order to make
several multiplications and additions before performing the division
(which is then delayed), even with a small storage. Indeed, 
one needs to perform a division only when the intermediate
result is able to overflow. 

\medskip\par\noindent\begin{minipage}{13cm}
\hrule\vspace{3pt}
Blocked dot product
\vspace{3pt}\hrule\vspace{2pt}\begin{verbatim}
  res = 0;
  unsigned long i=0; if (K<DIM) while ( i < (DIM/K)*K ) {
      for(unsigned long j = 0; j < K; ++j, ++i) res += a[i]*b[i];
      res %= P;
  }
  for(; i< DIM; ++i) res += a[i]*b[i];
  res %= P;
\end{verbatim}\hrule\end{minipage}

\begin{figure}[htb]\begin{center}
\includegraphics[width=\columnwidth]{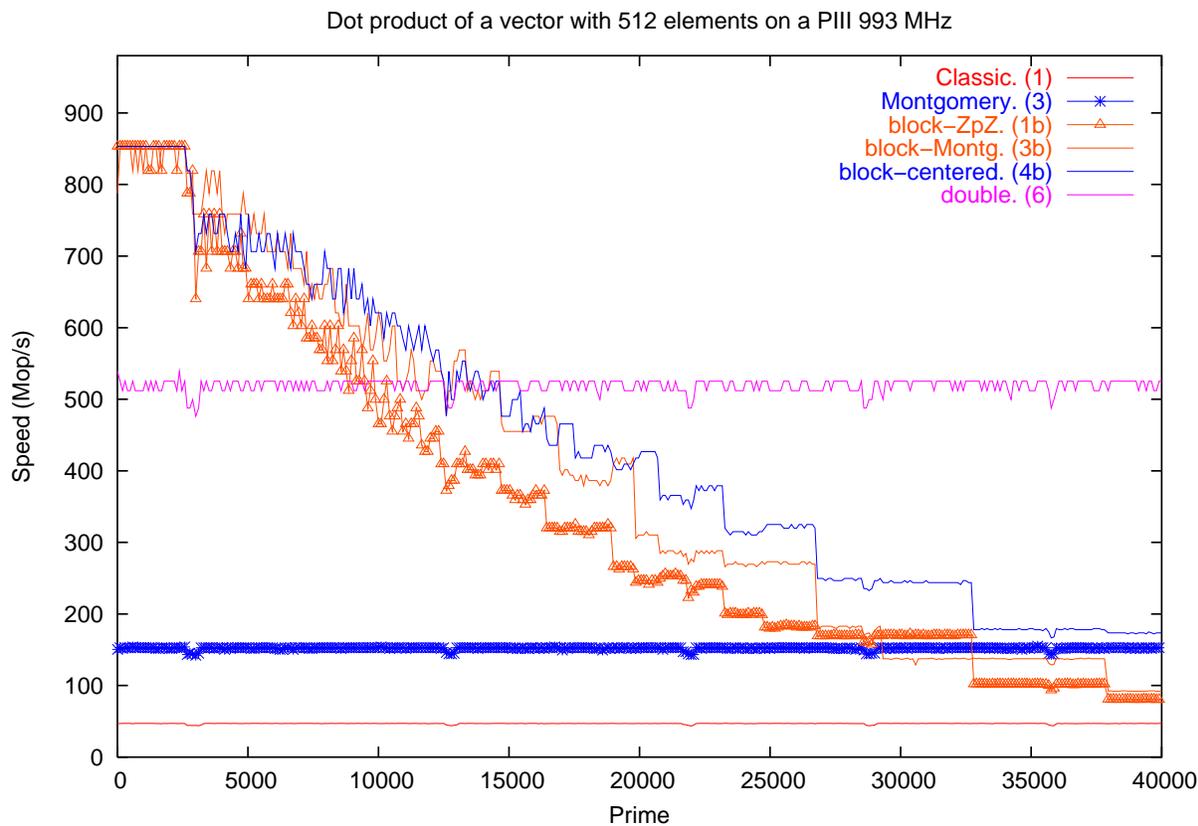}
\caption{Dot product by blocked and delayed division, on a PIII}
\label{fig:dotprod_Fbloc}
\end{center}\end{figure}

\bigskip
This method will be referred as ``block-XXX''. Figure
\ref{fig:dotprod_Fbloc} shows that this method is optimal for small
primes since it performs nearly one arithmetic operation per processor
cycle. Then, the step shape of the curve reflects the thresholds when
an additional division is made.

\subsubsection{Choice of Montgomery representation}\label{sssec:montg}
We see here, that our choice of representation ($aB$) for Montgomery
is interesting. Indeed, the basic dot product
operation is a cumulative AXPY. Now, each one of the
added values is in fact the result of a multiplication. Therefore,
the additions are between elements of the form $x_i B^2$, so that the
reduction can indeed be delayed.

\subsubsection{Centered representation}\label{sssec:center}
Another idea is to use a centered representation for ``Z/pZ'': indeed
if elements are stored between $-\frac{p-1}{2}$ and $\frac{p-1}{2}$, one can double the
sizes of the blocks (equation \ref{eq:fond} now becomes
$\lambda_{centered} (\frac{p-1}{2})^2 < 2^{m-1}$).

\subsection{Division on demand}\label{ssec:ov}
The second idea is to let the
overflow occur ! Then one should detect this overflow and
correct the result if needed. Indeed, suppose that we have added a
product $ab$ to the accumulated result $t$ and that an overflow has
occurred. The variable $t$ now contains actually $t-2^m$. Well, the
idea is just to precompute a correction $CORR=2^m \bmod p$ and add this
correction whenever an overflow has occurred.
Now for the overflow detection, we use the following trick: 
since $0<ab<2^m$, an overflow has occurred
if and only if $t+ab < t$!
The ``Z/pZ'' code now should look like the following:
\medskip\par\noindent\begin{minipage}{13cm}
\hrule\vspace{3pt}
Overflow detection trick
\vspace{3pt}\hrule\vspace{2pt}\begin{verbatim}
   sum = 0;
   for(unsigned long i = 0; i<DIM; ++i) {
         product = a[i]*b[i];
         sum += product;
         if (sum < product) sum += CORR;
   }
\end{verbatim}\hrule\end{minipage}

\bigskip
Of course one can also apply this trick to Montgomery reduction, but
as shown in figure \ref{fig:dotprod_Fbloc} the correction now implies
two reductions at the end. Indeed the trick of using a representation 
storing $aB \mod p$ for any element $a$ enables to perform only one
reduction at the end of each block. However at the end of the dot
product, one reduction has to be performed to mod out the result, and an
additional one to divide the element by $B$. Therefore, even though
Montgomery reduction is slightly faster than machine remaindering,
``overflow-Montgomery'' performs $k+1$ reductions when ``Z/pZ''
performs only $k$ machine remaindering.
\begin{figure}[htb]\begin{center}
\includegraphics[width=\columnwidth]{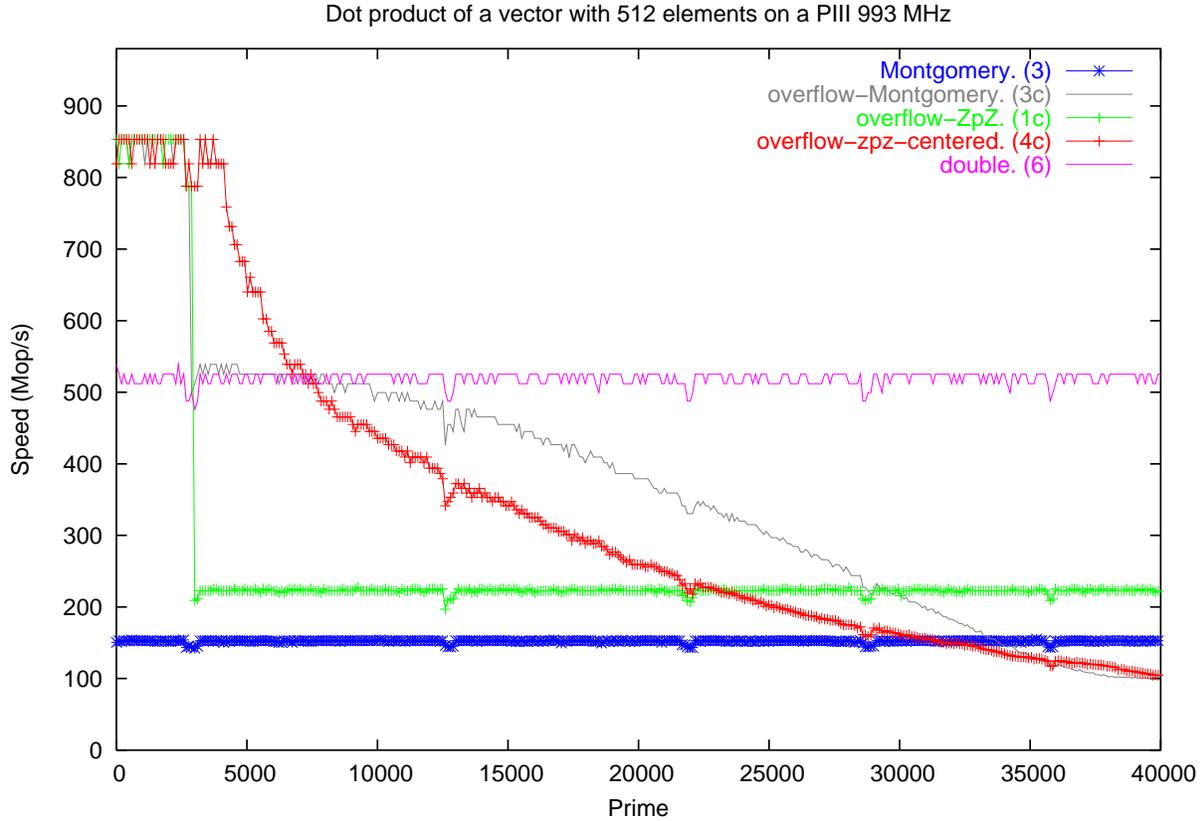}
\caption{Overflow detection, on a PIII}
\label{fig:dotprod_Fovfl}
\end{center}\end{figure}

The centered representation can be used also in this case.
This gives the better performances for small primes. 
However,
for this representation, the unsigned trick does not apply directly
anymore. The overflow and underflow need to be detected each by two tests:
\medskip\par\noindent\begin{minipage}{13cm}
\hrule\vspace{3pt}
Signed overflow detection trick
\vspace{3pt}\hrule\vspace{2pt}\begin{verbatim}
         if      ((sum < product) && (product - sum < 0)) sum += CORR;
         else if ((product < sum) && (sum - product < 0)) sum -= CORR;
   }
\end{verbatim}\hrule\end{minipage}

\bigskip
Thus, the total of four tests is  costlier and for bigger primes this
overhead is too expensive.

\subsection{Hybrid}
Of course, one can mix both \ref{ssec:del} and \ref{ssec:ov} approaches and delay even the overflow
test when $p$ is small. 
One has just to slightly change the bound on $\lambda$ so that, 
when adding the correction, no new overflow occur: 
\begin{equation}
\lambda(p-1)^2+(p-1)= \lambda p(p-1) < 2^m .
\end{equation}
This method will be referred as ``block-overflow-XXX''.\\
We compared those ideas on a vector of size
$512$ with $32$ bits (\texttt{unsigned long} on PIII). 
First, we see that as long as $512 p(p-1)<2^m$,
we obtain {\em quasi
optimal performances}, since only one division is performed at the
end. Then, when the prime exceeds the bound (i.e. for
$p(p-1)>2^{32-9}$, which is $p>2897$) an extra division has to
be made. On the one hand, this is dramatically shown by the drops of figure
\ref{fig:dotprod_Fovfl}. \\

\begin{figure}[htbp]\begin{center}
\includegraphics[width=\columnwidth]{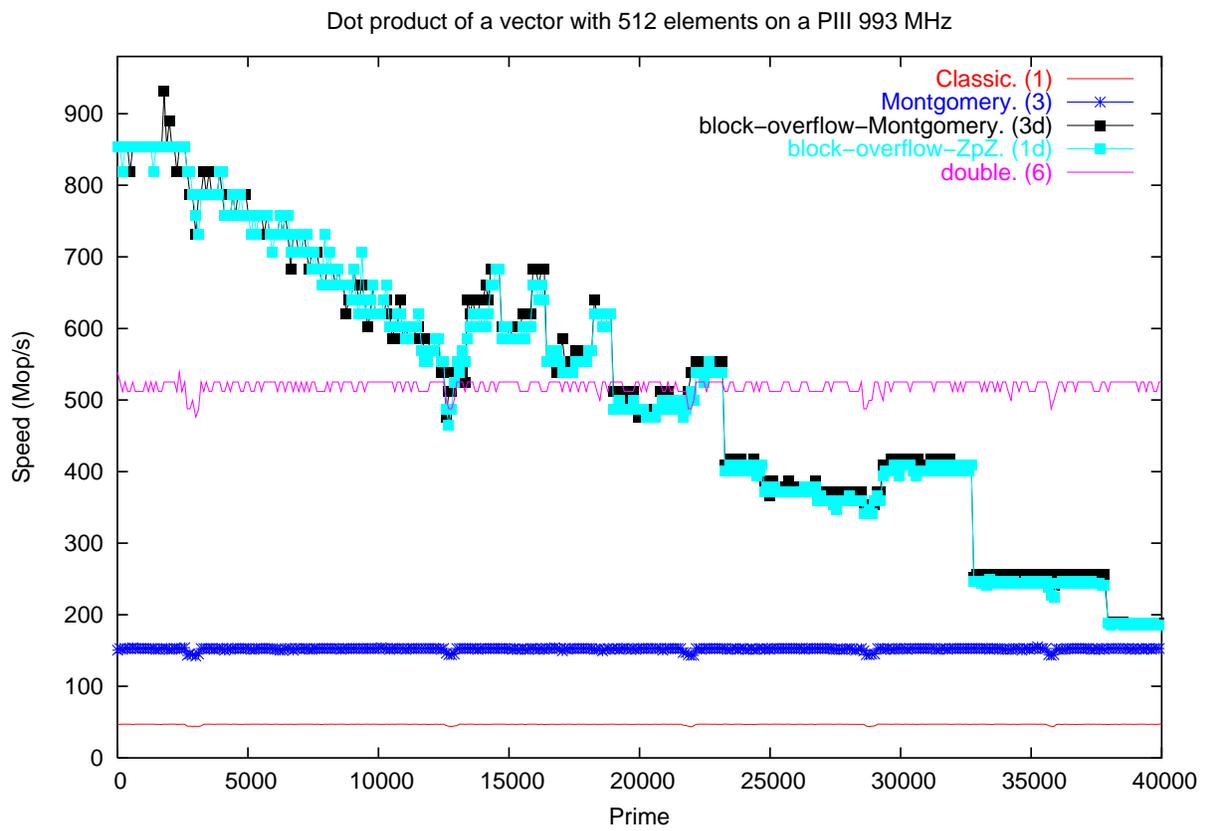}
\caption{Block and overflow, on a PIII}
\label{fig:dotprod_Fhyb}
\end{center}\end{figure}

On the other hand, however, those drops are levelled by our
hybrid approach as shown in figure \ref{fig:dotprod_Fhyb}.
There the step form
of the curve shows that every time a supplementary division is added,
performances drops accordingly.
Now, as the prime size augments, the block management overhead becomes
too important. \\

Unfortunately, an hybrid {\em centered} version is not useful. Even though
the overflow-centered was the best of the overflow methods, a block
version is of no use since the correction does not overflow towards
zero. After the signed overflow trick (resp. underflow), the current sum is of value
close to $-\frac{p-1}{2}$ or to $\frac{p-1}{2}$. Therefore no blocking
can be made since a single new product in the bad direction would
now induce an underflow (resp. overflow) ! \\

Lastly, one can remark than no significant difference exist between
the performances of ``block-overflow-Zpz'' and
``block-overflow-Montgomery''. Indeed, the code is now exactly the
same except for a single reduction for the whole dot product. This
makes the Montgomery version better but only extremely slightly.


\newpage
\section{Conclusion}
\begin{figure}[htbp]\begin{center}
\includegraphics[width=\columnwidth]{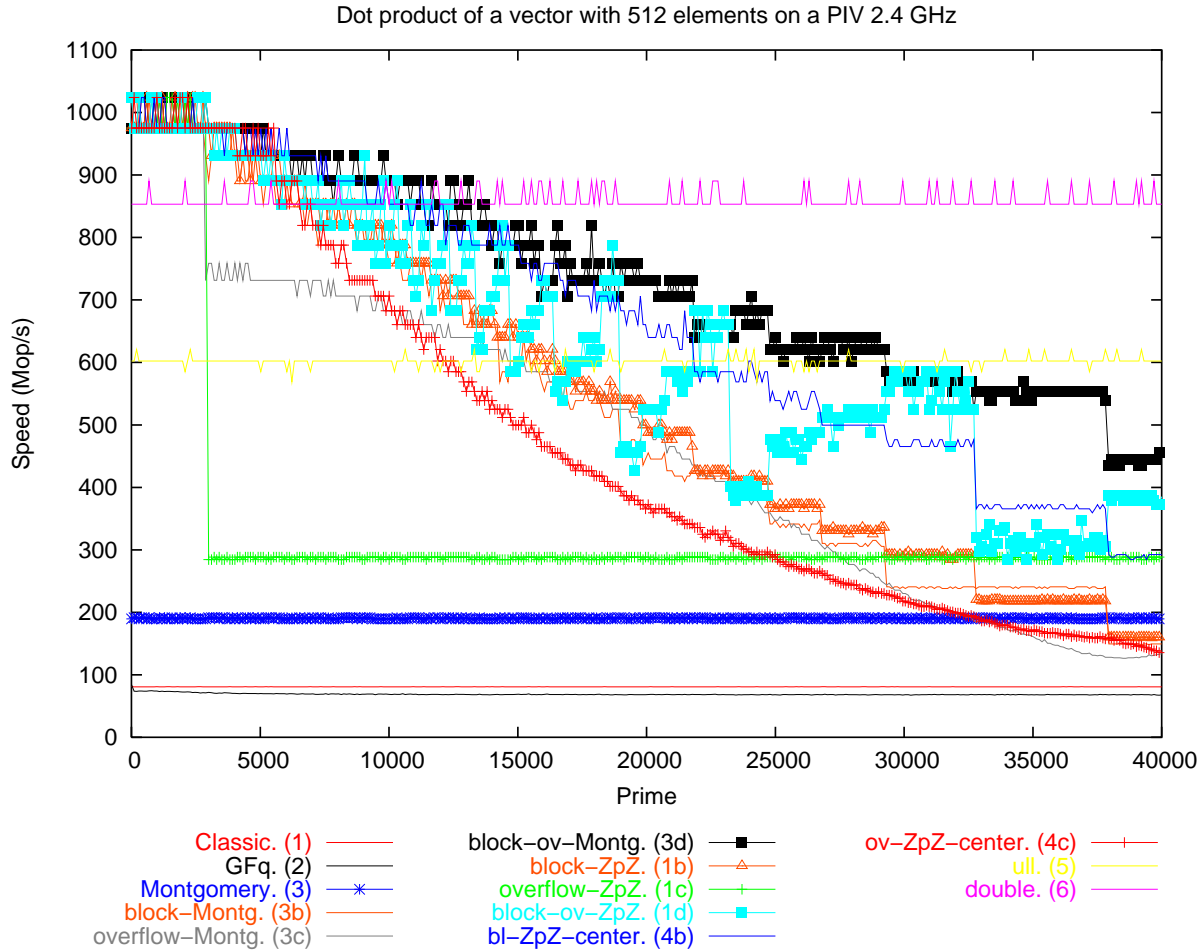}
\caption{Hybrid, on a PIV 2.4 GHz}
\label{fig:dotprod_PIV}
\end{center}\end{figure}

We have seen different ways to implement a dot product over word size
finite fields. The conclusion is that most of the times, a floating
point representation is the best implementation. This is even more the
case for a Pentium IV 2.4 GHz, as shown figure \ref{fig:dotprod_PIV}. 
However, with some care, it is possible to improve this speed for
small primes by a hybrid method using overflow detection and delayed
division. Now, the floating point representation approximately 
doubles its performances on the PIV 2.4 GHz, when compared to the 1 GHz
Pentium III. But surprisingly, the hybrid versions only slightly
improve. Still and all, an optimal version should switch from
block methods to a floating point representation according to the vector
and prime size and to the architecture.\\
 
Nevertheless, 
bases for the construction of an optimal dot product over word size
finite fields now exist: the idea
is to use an Automated Empirical Optimization of Software
\cite{Whaley:2001:AEO} in order to produce a library which would
determine and choose the best
switching thresholds at install time.

\section*{Acknowledgements}
Grateful thanks to Paul Zimmermann for invaluable advice and comments
and several implementations.

\bibliographystyle{plain}
\bibliography{dotprod}

\end{document}